# Self-navigation of STM tip toward a micron sized sample

Guohong Li, Adina Luican, and Eva Y. Andrei

*Department of Physics & Astronomy, Rutgers University, Piscataway,*

*New Jersey 08854, USA*

We demonstrate a simple capacitive based method to quickly and efficiently locate micron size conductive samples on insulating substrates in a scanning tunneling microscope (STM). By using edge recognition the method is designed to locate and identify small features when the STM tip is far above the surface allowing for crash-free search and navigation. The method can be implemented in any STM environment even at low temperatures and in strong magnetic field, with minimal or no hardware modifications.

## I. INTRODUCTION

A scanning tunneling microscope[1,2] (STM) is a powerful tool to study materials with atomic resolution. In the STM topography mode a sharp metallic tip scans above a conductive sample surface while monitoring the tip-sample tunneling current which depends exponentially on the distance. Typically the measurement is carried out at a tip-sample distance of order 1 nm. The precise control of tip position (x, y, z) is usually realized by employing a piezo-electric tube scanner with pico-meter resolution[2]. However, the high resolution of STM sets a limit to its field of view, the largest of which is usually about 1μm.

In practice[2], one needs additional coarse positioning stages for tip navigation parallel to the sample surface (x, y) and for perpendicular (z) approach. These stages typically travel distances of a few mm with step size of 10-1000 nm. Although a reliable z-stage is a must for any STM, the x-y navigation is not crucial for standard STM measurements on mm sized samples because the typical features appear everywhere.

Recently there has been increased interest in studying micron sized samples, such as a graphene flake on a $SiO_2$ substrate. However, because the STM is intrinsically nearsighted it is quite challenging to locate a specific micron sized sample on a macroscopic substrate. Since STM images topography it is natural to try using topographic features[3,4] as guides. But this method is not practical because of the limitations inherent to the technique: the small field of view requires comparing many images frame by frame; the even smaller working distance does not favor high speed scanning due to the risk of tip crashing; the finite dynamic range of piezo-electric scanner in z direction, typically less than 300nm, usually cannot cover the height variation over distances of tens of µm and therefore frequent tip retracting and approaching are necessary during the navigation. These factors render the STM navigation extremely slow. A further complication arises in the presence of insulating-conductive-boundaries, where navigating the STM tip will surely result in a crash into nearby insulating areas.

It is possible to circumvent these complications by using additional setups, such as a scanning electron microscope[5] or a long range optical microscope[6], to visualize the STM tip during navigation. However these external aids are impractical at low temperatures and in high magnetic fields due to the harsh environment and limited space.

Here we report a capacitance-based method to navigate the STM tip, which allows finding micron sized samples quickly and efficiently. The method consists of back-gate compensation[7,8], refocusing during navigation, and distinguishing edges of conducting electrodes and the sample. It requires no additional setup other than two independent bias voltages, one for the sample and the other for a back-gate. The latter is simply a conductive plane below the sample separated by a thin insulating layer.

## II. CAPACITIVE PICKUP AND COMPENSATION

The signal of interest in STM is the tunneling current:

$$I = G_t V_s \tag{1}$$

where $V_s$ is the sample bias voltage and $G_t$ the tunneling conductance. To measure scanning tunneling spectroscopy[2], one usually applies a small AC modulation, $\tilde{V}_s$, to the sample bias voltage so that there is an AC current, $\tilde{I}$, flowing through the STM tip,

$$\tilde{I} = G_t \tilde{V}_s + i\omega C \tilde{V}_s \tag{2}$$

There are two contributions to the AC current, one from tunneling (first term) and the other from capacitive pickup (second term) via the effective tip-sample capacitance $C$. Since the capacitive current is not limited to the tunneling regime, the two contributions are easily distinguishable. The pickup current is a function of the geometry through the tip-sample capacitance and can be used to roughly monitor tip approach starting from a few mm away from the sample surface. We will show that this signal can be used to resolve small structures.

In principle, capacitive currents should depend on the relative lateral (x and y) position between tip and sample. As we observed experimentally, the effective capacitance change is ~30 $aF$ ($10^{-18}F$) when an STM tip with 3μm tip-sample distance scans across a 5 μm conductive strip deposited on an insulating substrate. However, the background pickup is usually several orders of magnitude larger, ~ 6 fF ($10^{-15}F$). To measure the small change in capacitance, one needs very good background compensation, which could be accomplished with a capacitance bridge setup. We show that in most STM designs it is in fact possible, with only a minor modification to use the sample as part of a built-in bridge circuit. For example, a graphene flake on $SiO_2$ is supported on a heavily doped silicon gate electrode. This gate can act as one arm of the capacitance bridge as illustrated in Fig. 1(b) and thus one can tune the out-of-phase voltage -$\tilde{V}_{gate}$ to cancel the background current. In this circuit the variation of the tip current, $d\tilde{I}$, is directly proportional to $dC_{t-s}$, the tip-sample capacitance and so are their spatial derivatives $\frac{d\tilde{I}}{dx} \propto \frac{dC_{t-s}}{dx}$.

It is worth noting that the presence of back gate also changes the electric field distribution around the sample. Without the gate as shown in Fig.1(c), the equipotential lines quickly lose the shape of the sample. A grounded back gate pushes up the equipotential lines [Fig.1(d)]. However, a dramatic change occurs when an opposite voltage is applied to the back gate. As seen in Fig.1(e), the equipotential lines develop strongly varying features above the edges which remain sharp and clearly distinguishable at distances comparable to the sample size. Thus one could resolve the thin bar with large tip sample distance. We next focus on the experimental results.

Although the method reported here was successfully used in our recent study[7] of Landau levels in graphene on $SiO_2$ at 4.4K and 12T, the data reported here were taken at room temperature in order to have an independent check on tip navigation from an optical microscope. Typical AC voltages applied to the sample were 200 mV$_{rms}$ at 5 kHz. The frequency is limited by the bandwidth of the current amplifier ($10^9$V/A gain), which is also used to measure tunneling currents.

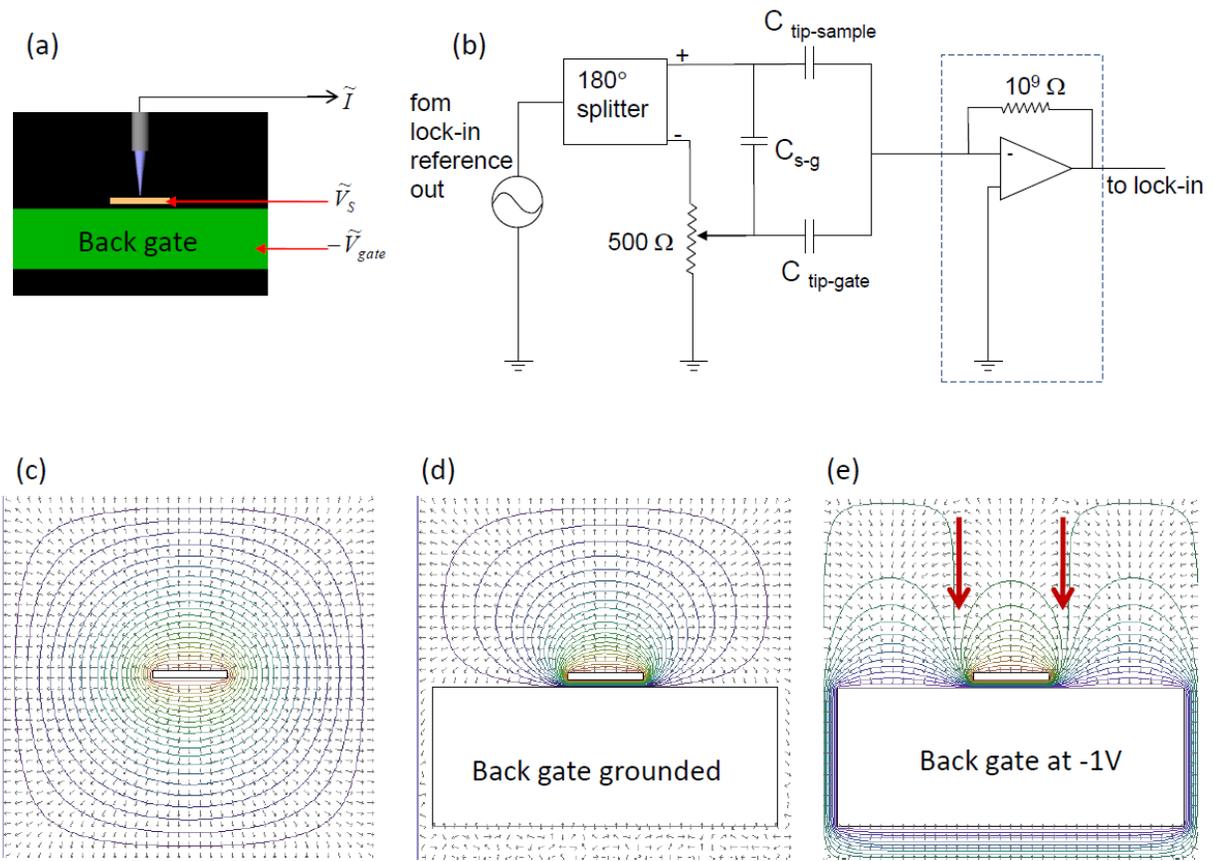

FIG.1. (Color online) Use of back gate in capacitance measurements. (a) Schematic experimental setup. (b) Equivalent circuit. The output voltage from the reference channel of a lock-in amplifier is split into two with 180° phase shift. One (+) is applied to the sample directly as $\tilde{V}_s$, the other (-) is applied to the gate ($-\tilde{V}_{gate}$) through a pot resistor to adjust the amplitude. Capacitive pickup currents are measured with the same amplifier (hashed area) that is used for tunneling currents. (c) Electric field distribution near a conducting bar extending out of the paper. The strip is at 1V. Lines are equipotential contours and arrows show field directions. (d) and (e) Same as (c) but with a nearby back gate grounded and at -1V, respectively. Red arrows mark the steep potential lines near the sample edges. (c) to (e) were calculated with *Field Precision TC* (educational 7.0).

## III. EDGE DETECTION

Fig.2 shows the variation of the capacitive currents as the STM tip moves above a 200 μm wide Au film. The tip is 60 μm away from the sample surface in Fig.2(a). As the tip moves across the sample, the current peaks near the center of the film. The asymmetry is due to the background, e.g. overall sample geometry and wiring. This background current could significantly distort the peak when the tip is 210 μm away from the sample surface, as shown in Fig.2(b).

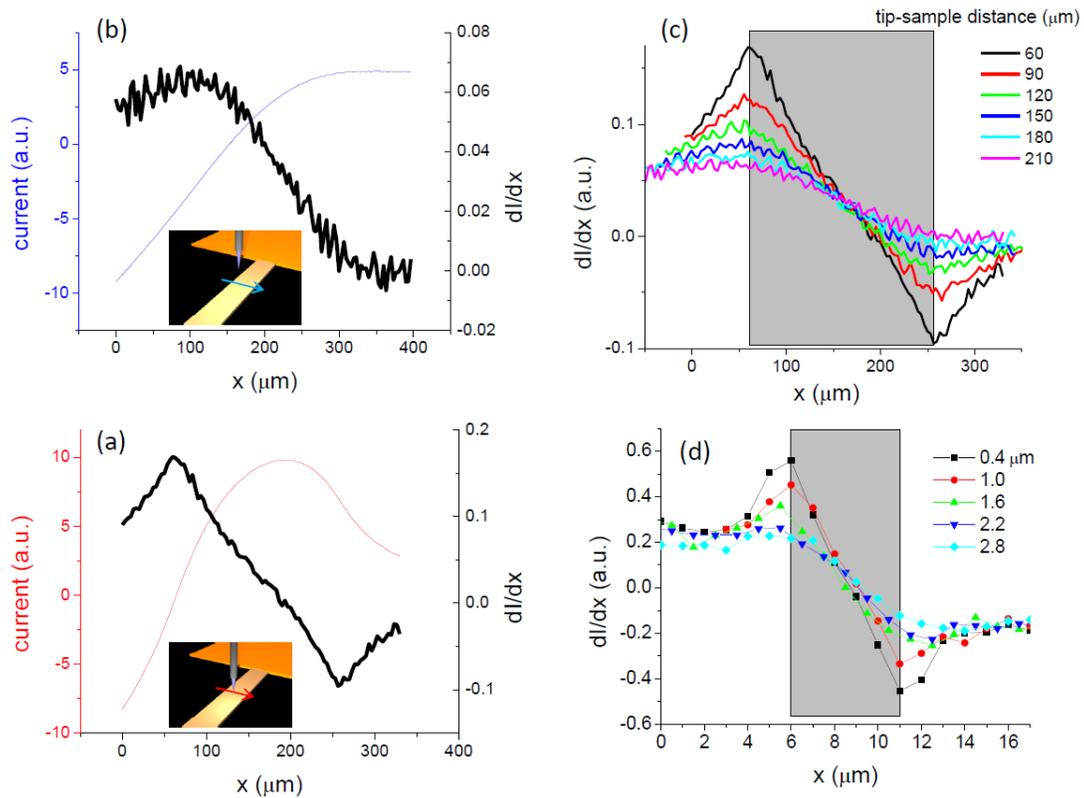

FIG.2. (Color online) Variation of capacitive currents as STM tip moves above thin film. (a) Tip 60 μm above a 200 μm wide film. Thin line: pickup current. Thick line: spatial derivative dI/dx. Insert: schematic of tip sample geometry. (b) Same as (a) but with tip-sample distance of 210 μm. (c) spatial derivative dI/dx for various tip-sample distances. Data are centered relative to the shaded area that shows the 200 μm wide sample. Shaded area marks the film width. (d) Same as (c) but for a film with 5μm width.

Although the capacitive currents in Figs. 2(a) and (b) look quite different, their spatial derivatives, *dI/dx*, appear to be similar and symmetric. The two turning points in *dI/dx* correspond to the edges of the film. Fig. 2(c) plots a systematic study for various tip sample distances. The sharpness of the turning points decreases with increasing distance. Still the central part of the sample can be easily identified even if the tip-sample distance is comparable to the sample width. However, to identify the edges accurately, the distance has to be sufficiently small compared to the sample width so as to produce sharp turning points above the sample edges. For example, in Fig.2(d), the tip-sample distance has to be less than ~1.6μm to resolve the edges of a 5μm wide film.

To check the reliability of this method, we repeated the measurements on films with different width shown in Fig. 3(b). As illustrated in Figs. 2(c) and 2(d), the *dI/dx* signal exhibits sharp peaks at the position of the sample edges if the "aspect ratio" (tip height over sample width) is less than ~0.3. Therefore as the features become smaller it is important to approach the tip toward the sample surface in order to obtain the desired resolution. This is similar to the concomitant focus and magnification adjustment in an optical microscope.

Fig. 3(a) compares the measured pad size obtained by this method with the actual size obtained from the optical image in Fig. 3(b). The measured sizes are very close to the actual ones for all film width down to the smallest measured feature provided the aspect ratio is kept below 0.3. For larger aspect ratios the measured widths are systematically larger than the actual widths. This error if not taken into account could lead to a fatal tip crash. Below we describe a navigation protocol which allows finding a micron-size sample reliably and safely.

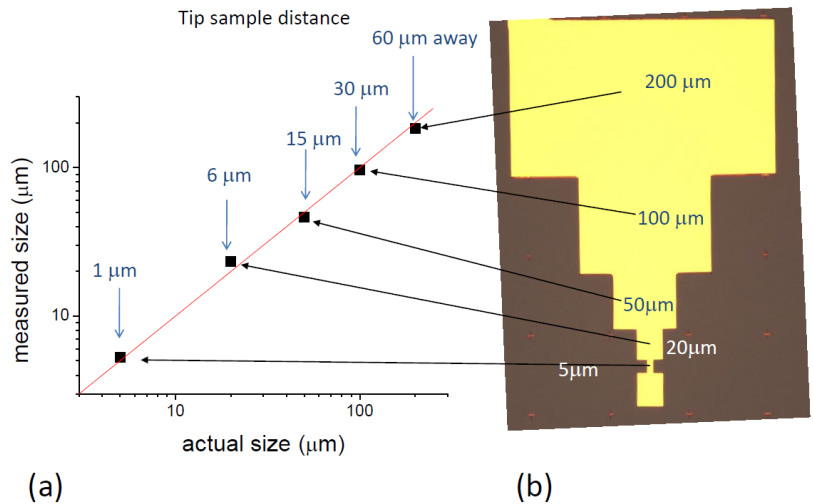

FIG. 3. (Color on line) Sample widths measured by the capacitance method illustrated in Fig.2. (a) Measured size versus actual size. Tip sample distances adjusted for sharp edge resolution corresponding to aspect ratios of 0.3 or less. Red line is a guide to the eye through the values of the measured widths. (b) Optical image of sample.

## IV. PROTOCOL OF NAVIGATION

Before demonstrating how to find a micron sized sample, we emphasize the key points in the navigation protocol.

a) Identify the central region of a conductive sample from the spatial derivative of the capacitive currents. This step, which is done with the tip far away from the sample surface, (Fig.2) guarantees that the tip is targeting a conductive region.

b) Use the STM mode to find the sample surface safely and retract the tip to a height corresponding to an aspect ratio of ~0.3. This step, similar to re-focusing in an optical microscope, enables sharper contrast.

c) Identify the edges of a conductive sample from the spatial derivative of capacitive currents.

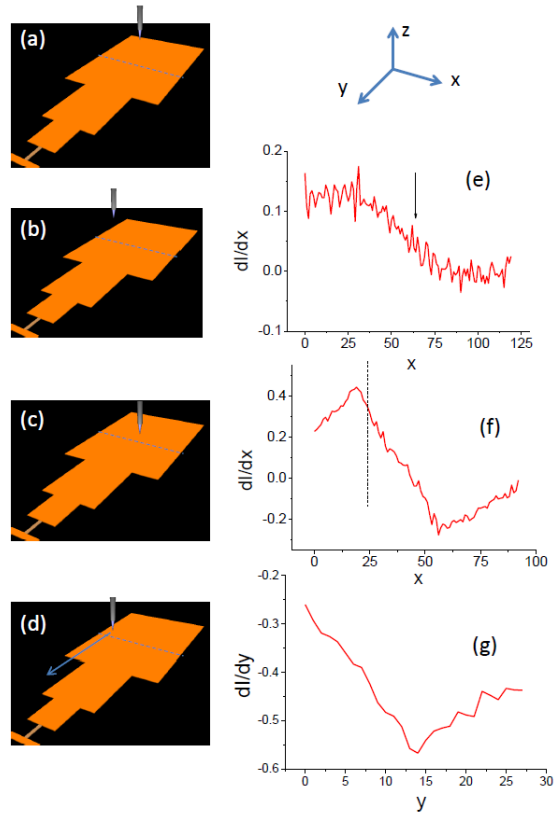

FIG.4. (Color online) Protocol of navigation. (a) Initial alignment between tip and sample at room temperature. (b) Possible drift of tip position after transferring and cooling down. (c) Tip position adjusted by centering and working in STM mode as discussed in text. (d) Tip near one edge ready to search for the lower boundary of the pad. (e) Derivative of capacitive currents in the first scan along the dashed line shown in (b). Arrow indentifies the center of the pad (f) Derivative of capacitive currents in along the dashed line in (c) with a smaller tip sample distance. Dotted line marks the choice of tip position in (d). (g) Derivative of capacitive currents in along the arrow in (d) to find the lower boundary. Units in (e)-(g) are arbitrary. Space directions are defined in the upper-right insert.

Following these key points, navigation procedure will depend on the geometry of the metal pads contacting the micron sized sample of interest. Below we show how this procedure is applied to locate the sample in the left-lower corner of Fig.4 (a).

Starting at room temperature, we can easily position the tip ~0.2mm above the biggest pad using optical access [Fig.4(a)]. Once the STM is transferred into a magnet and is cooled down to 4K, the tip could drift away [Fig.4(b)] because of mechanical perturbations and thermal contractions. However, with a well-designed STM head, the drift is rarely larger than 100 μm in any direction.

Considering the weak tip height dependence of the overall structure in dI/dx (Fig.2), one can always do an initial scan across the big pad to identify its center from dI/dx in Fig.4(e) and position the tip accordingly. Once it is established that the STM tip is targeting a conductive surface, the surface is found using the STM tip-approach mechanism. During this stage the large AC modulation is turned off and the DC sample bias voltage, say 500mV, remains. When the surface is found, the tip is retracted to the appropriate distance [Fig.4.(c)]. This distance should be large enough to accommodate height variations of the sample during tip movements. We note that the height variation does not affect the identification of edges significantly because of the weak dependence shown in Fig. 2(c).

With reduced tip-sample distance, the edges can be identified with better accuracy as shown in Fig. 4(f). Subsequently the tip is positioned near one of the edges [Fig.4(d)] and moved along the edge. The lower boundary of the big pad appears as a dip in dI/dy [Fig.4(g)].

The procedure is then repeated on the smaller pad. For optimal imaging conditions the tip is retracted less after finding the surface of the smaller pad so as to maintain the aspect ratio within 0.3. Thus, as the tip approaches the targeted sample it also gets closer to the surface.

## V. CONCLUSIONS

We have demonstrated a capacitive pickup method and a navigation protocol to locate small samples on an insulating substrate using an STM tip. The method employs metallic guiding pads whose edges can be clearly identified by using back gate compensation and the spatial derivative of capacitive currents. This capacitance based method involves minimal modifications to the STM setup and can be applied more generally to other scanning probe microscopes equipped with coarse motor navigation.

**ACKNOWLEDGEMENTS**

This work was supported by the DOE under Grant No. DE-FG02-99ER45742 and partially supported by Grant No. NSF-DMR-0906711 and by Lucent.